\documentclass[epj]{svjour}
\pdfoutput=1

\usepackage{graphicx}

\begin{document}

\title{A complex network approach to cloud computing}

\author{Gonzalo Travieso 
  \and Carlos Ant\^onio Ruggiero 
  \and Odemir Martinez Bruno 
  \and Luciano da Fontoura Costa}
\institute{Instituto de F\'isica de S\~ao Carlos, Universidade de
  S\~ao Paulo, S\~ao Carlos, SP, Brazil\\
\email{gonzalo@ifsc.usp.br}}

\date{}

\abstract{ Cloud computing has become an important means to speed up
  computing.  One problem influencing heavily the performance of such
  systems is the choice of nodes as servers responsible for executing
  the users' tasks.  In this article we report how complex networks
  can be used to model such a problem.  More specifically, we
  investigate the performance of the processing respectively to cloud
  systems underlain by Erd\H{o}s-R\'enyi and Barab\'asi-Albert
  topology containing two servers.  Cloud networks involving two
  communities not necessarily of the same size are also considered in
  our analysis.  The performance of each configuration is quantified
  in terms of two indices: the \emph{cost} of communication between
  the user and the nearest server, and the \emph{balance} of the
  distribution of tasks between the two servers.  Regarding the latter
  index, the ER topology provides better performance than the BA case
  for smaller average degrees and opposite behavior for larger average
  degrees.  With respect to the cost, smaller values are found in the
  BA topology irrespective of the average degree.  In addition, we
  also verified that it is easier to find good servers in the ER than
  in BA.\@ Surprisingly, balance and cost are not too much affected by
  the presence of communities.  However, for a well-defined community
  network, we found that it is important to assign each server to a
  different community so as to achieve better performance.   }

\maketitle

\section{Introduction}

With the booming of the Internet, an impressive mass of computing
resources, encompassing both machine and data, became broadly
available.  At the same time, the number of users grew largely,
implying in a growing demand to Internet collaborative access in a
number of machines and platforms.  Cloud computing emerged as the
natural integration of these two trends.  The basic idea in this
paradigm is to define integrated, distributed, servers capable of
supplying services to users through Internet.  In addition, since the
data in the cloud system has to be widely accessible in many places
and for many users, multiple servers are required.  As a consequence
of its reliance on the Internet, cloud systems tend to have complex
topologies, which compounds the choice of where in the network the
servers should be placed.  In particular, the distribution of the
servers should lead to small communication times between users and
servers, without overloading any of the servers.

Complex networks have become an important subject in science and
technology because of their ability to represent and model a large
number of complex systems such as society, protein interaction,
transportation, among many
others~\cite{Barabasi:2002:LIN:829575,Newman:2010:NI:1809753,costa2007characterization}.
In computer science, complex networks have been used, for instance, in
the study of the topology of the Internet~\cite{faloutsos99:_inter},
the Web~\cite{baldi03:_model_inter_web}, email
communications~\cite{tyler03:_email}, the complexity of software
systems~\cite{ma05}, and modeling grid
computing˜\cite{da2005complex,de2008effects,travieso2011effective,travieso2013predicting,prieto-castillo14}.
In the latter field, complex networks were used to represent task
execution in grid computing environments, with the tasks being
supplied by a master, on demand from worker processors, which were
distributed along the network topology.  Contrariwise, in cloud
computing several users concur for access to a small number of
servers.

In the present work we extend the use of complex networks to modeling
and evaluating the performance of multiple-severs cloud computing
environments.  More specifically, we quantify the effect of different
topologies ---namely Erdos-R\'enyi˜\cite{Erdos-Renyi:1959},
Barabasi-Albert˜\cite{Barabasi97} and a modular model--- with respect
to the positioning of servers in the network topology.  For
simplicity's sake, we consider only pairs of servers in the cloud
environments.

The article starts by presenting the basic concepts and methods
adopted, and follows by presenting how cloud environments can be
represented in complex networks, and investigating the performance of
such configurations for different placements of servers in the network
topology.  We have found that the distribution of servers in cloud
computing environments is determinant for the performance, quantified
in terms of communication cost and balance.  In addition, the best
configurations depend strongly on the network topology.

\label{sec:intro}

\section{Methodology}
\label{sec:meth}

\begin{figure*}[tbp]
  \centering
  \includegraphics[width=0.95\textwidth]{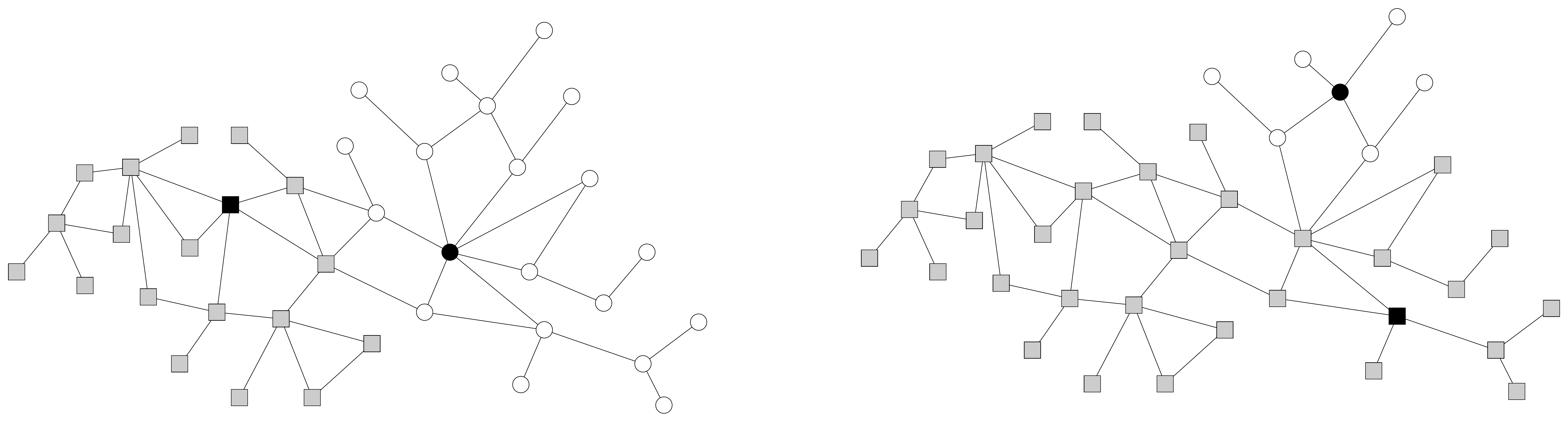}
  \caption{Distribution of clients to servers in a network. Each
    client is associated with the closest server (using shortest path
    distances). On the left, the two servers (marked with a dark
    color) are responsible for the same number of clients; on the
    right, the server choice results in an inbalanced distribution.}
  \label{fig:dist}
\end{figure*}

We consider here a network that provides a communication
infrastructure for agents placed on its nodes.  Some agents (called
``servers'') will be chosen to provide services for the other agents
(the ``clients'').  A request for a service is forwarded by a client
to the closest server following a shortest path, and the response from
the server follows the same path in the reverse order.  Once the
servers are placed in the network, each client is assigned to the
closest server.  Thus, for good efficiency on the delivery and
execution of the services, the servers must be placed in the network
such that they are relatively close to the their clients and each
server is responsible for answering requests from about the same
number of other clients.  Figure~\ref{fig:dist} shows two contrasting
situations regarding the placement of two servers in a same network.
On the left part of the figure, a good balance is achieved because
each server is associated to similar number of clients.  Contrariwise,
on the right, one of the servers resulted with only six clients, while
a much larger number of clients is associated to the other server.

To quantitatively evaluate the aspects above, given a choice of servers we
compute two measurement: the \emph{average cost} and the
\emph{balance}, defined as follows.

Let $s(i)$ be the server associated with client $i$ (i.e.\ the server
that is closest to $i$).  The average cost is defined as
\begin{equation}
  \label{eq:cost}
  c = 2 \sum_i d(i, s(i)),
\end{equation}
where $d(i,j)$ is the shortest path length from node $i$ to node
$j$ in the network.  The factor $2$ is include to account for the
request and response communication costs. The sum runs over all
clients $i$.

The balance should quantify if all servers receive work from
approximately the same number of clients.  Let $\mathcal{A}_j$ be the
set of clients associated with server $j$, and $|\mathcal{A}_j|$ its
cardinality.  We define the balance as the ratio from the smaller to
the larger of these sets:
\begin{equation}
  \label{eq:balance}
  b = \frac{\min_j |\mathcal{A}_j|}{\max_j |\mathcal{A}_j|},
\end{equation}
where the $\min$ and $\max$ run over all servers $j$.

We want to evaluate the effect of network topology on this dynamical
process.  For simplicity and computational efficiency, we consider the
case of only two servers.  Given a pair of servers, we choose to which
server the clients are associated, using the distance matrix of the
network and choosing the nearest server for each client.  Afterward
the values of average cost and balance are computed for this pair of
servers using the expressions above.  The process is repeated for all
pairs of servers in the network.  A good pair of servers should have
simultaneously a large value of balance and a small value of average
cost.  We define the \emph{elite} of server pairs as the intersection
of the pairs within the 20\% with best (smallest) values of average
cost and the 20\% with the best (largest) value of balance.  For each
evaluated network we compute: the smallest values of average cost and
largest value of balance for all pairs, the threshold values of
average cost and balance needed to include a pair in the elite, the
average values of average cost and balance for all pairs, the average
values of average cost and balance for the pairs in the elite, and the
number of pairs in the elite.

We consider now the effect of community structure in the network over
the balance and average cost.  We want to quantify the effects of
community separation and differences in community sizes.  The network
model used consists of $N$ nodes, each associated with a given
community.  We fix the number of communities in two, and associate
$[\alpha N]$ nodes to the first community and $N-[\alpha N]$ nodes to
the second, where $0<\alpha<1$ and $[x]$ means rounding $x$ to the
closest integer.  Without losing generality, in the following we
choose community 1 to be the smallest, and therefore $0<\alpha\le
1/2$.  Each pair of nodes is connected according with the following:
\begin{description}
\item[Inside community 1] If both nodes are from community 1, they are
  connected with probability
  \begin{equation}
    \label{eq:p1}
    p_1 = \frac{1-\delta}{\alpha}\frac{\left< k \right>}{N},
  \end{equation}
  where $\left< k \right>$ is the desired average degree and the
  parameter $\delta$ controls the community structure as will be
  discussed below.
\item[Inside community 2] When both nodes are from community 2, the
  connection probability is
  \begin{equation}
    \label{eq:p2}
    p_2 = \frac{1-\alpha-\alpha \delta}{(1-\alpha)^2}\frac{\left< k
      \right>}{N}.
  \end{equation}
\item[Between communities] If the nodes are in different communities,
  the probability of connection is given by
  \begin{equation}
    \label{eq:pi}
    p_i = \frac{\delta}{1-\alpha}\frac{\left< k \right>}{N}.
  \end{equation}
\end{description}
Different values are chosen for the probability in the two communities
to achieve the same average degrees for nodes in both communities.  If
the same value of probability is used for a small and a large
community, each node in the smaller community have less other possible
nodes inside the same community to connect, and therefore has a
smaller expected degrees than the nodes in the larger community.  For
the values in Equations~(\ref{eq:p1}) to~(\ref{eq:pi}), the average
degree of nodes in the first community is (for large values of $N$):
$$
p_1 \alpha N + p_i (1-\alpha) N = (1-\delta)\left< k \right>  +
\delta \left< k \right>  = \left< k \right>.
$$
For the second community, the average degree is:
$$
p_2 (1-\alpha) N + p_i \alpha N = \frac{1-\alpha-\alpha
  \delta}{1-\alpha}\left< k \right>  + \frac{\alpha \delta}{1-\alpha}
\left< k \right> = \left< k \right>.
$$

The value $\delta$ is a community strength parameter and quantifies
how much of the existing connectivity in the first community is used
for connections with the other community.  Note that, if $\delta=0$,
then $p_i = 0$ and there are no connections between the two
communities.  Therefore values of $\delta$ near zero result in a
pronounced community structure.  On the other hand, if $\delta=1$, we
have $p_1=0$, and all links from community~1 are to community~2.  In
this last case, if the two communities are of the same size
($\alpha=1/2$), all links from nodes in community~2 go to nodes in
community~1, and the network is bipartite.  In the general case, links
still exist among the nodes in the largest community.  A value of
$\delta=1/2$ corresponds to the case where half of the links in
community~1 go to the same community, and half to the other community
and is the largest value of interest to us here.

\section{Results and discussion}
\label{sec:res}

\subsection{ER and BA networks}

\begin{figure*}[tbp]
  \centering
  \includegraphics[height=0.9\textheight]{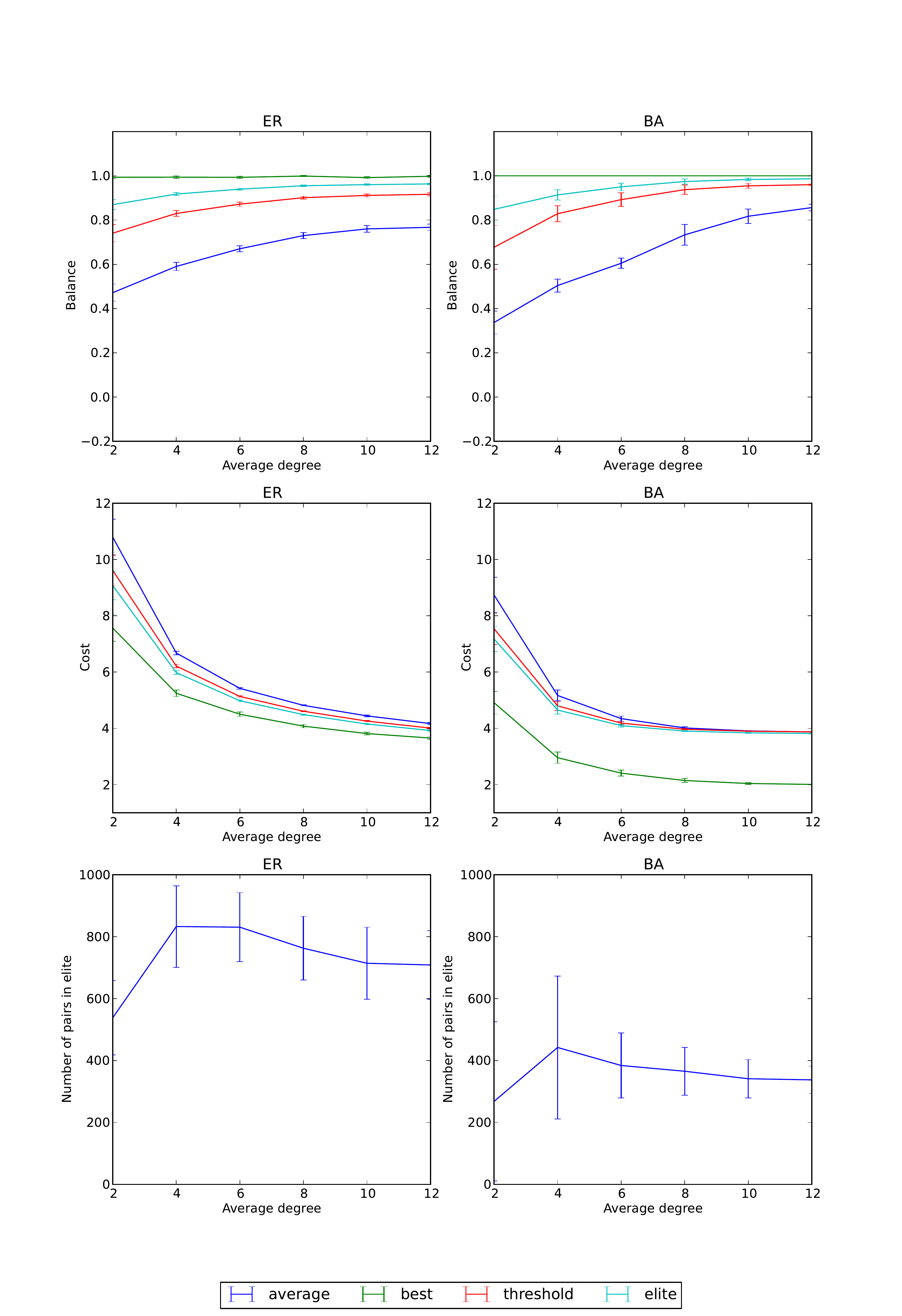}
  \caption{Balance, cost and number of pairs in the elite for BA and
    ER models.  Points are averages of 100~networks, each with
    200~nodes.  Error bars show one standard deviation.  All pairs of
    each network are evaluated.} \label{fig:cost-balance-elite}
\end{figure*}

Figure~\ref{fig:cost-balance-elite} shows the result of this
evaluation for the Erd\H{o}s-R\'{e}nyi (ER) and Barab\'{a}si-Albert
(BA) network models with varying values of average degree.  We used
these models to evaluate the effect of degree heterogeneity.  Each
network has $200$ nodes and we generate $100$ networks for each
model/parameter combination to compute average and standard deviation
of each measurement.

First we notice that the theoretical maximum value of balance is
achieved for some node pair in most networks.  Also, with the
exception of small values of average degree, the balance achieved by
pairs in the elite is close to the maximum in both models.  It can be
seen also that the values of balance (network and elite averages, as
well as threshold) for ER networks are slightly better than for BA
networks.  This is possibly due to the excessive influence of the hubs
in the BA topology, making it more sensitive to the choice of pairs.

The situation is different with regard to communication costs, where
BA networks are better (with the exception of network with high
average degree).  It is interesting to note also that the difference
in costs for the best and average pairs is much larger in BA networks.
This is due to the fact that in these networks, the hubs are central
(in the closeness centrality sense) and therefore have small average
distances to the other network nodes.  If two hubs are chosen in a
pair, the communication costs for the pair will be small.  But pairs
with two hubs are a small minority of all the possible pairs, and
therefore do not significantly affect the averages.

The networks have $N=200$ nodes, and therefore there are about
$N^2/2 = 20000$ distinct pair.  For the elite, we choose the pairs
that are in the 20\% better in cost and in balance.  If the two
criteria were unrelated, the expected number of pairs in the elite
would be $0.04 N^2/2 = 800$.  As can be seen in
Figure~\ref{fig:cost-balance-elite}, the number of pairs in the elite
of ER networks is close to this expected value, with significant
differences only for small average degrees.  On the other hand, in BA
networks the number of pairs in the elite is much lower, about half of
the number in the ER networks.  This suggests that in topologies with
strong degree heterogeneity the efficiency is much more sensitive to
the choice of the pair of servers.

\subsection{Communities}

\begin{figure*}[tbp]
  \centering
  \includegraphics[height=0.9\textheight]{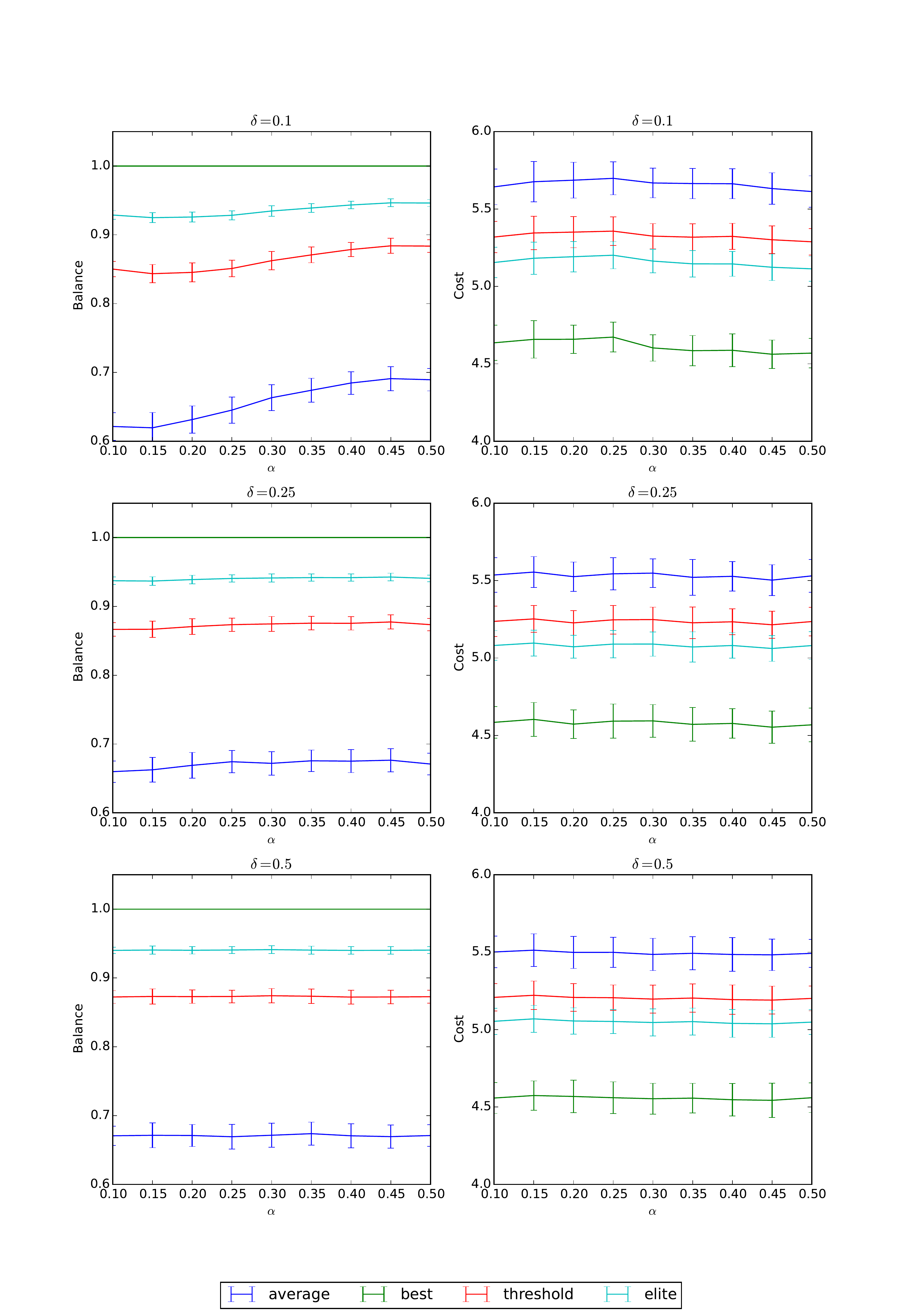}
  \caption{Balance and cost in a community model as a function of
    community sizes.  The network has two communities, with nodes
    distributed between them according to parameter $\alpha$ (values
    of $\alpha$ close to $1/2$ imply communities of similar size, see
    text).  The connectivity between nodes in the two networks is
    controled by parameter $\delta$, for which some values are chosen
    (larger values of $\delta$ imply more connections between
    communities, see text).  Results are averages of 100~networks,
    each with 200~nodes; error bars show one standard
    deviation.}\label{fig:comm-alpha}
\end{figure*}

Figure~\ref{fig:comm-alpha} shows the impact of changes in community
sizes ($\alpha$) in the balance and cost, for different values of
$\delta$. As expected, for $\delta=1/2$ there is no influence of the
division of nodes in communities, as the communities are not well
separated.  For smaller values of $\delta$ we can see some influence
of $\alpha$ in the balance, but almost no influence in cost.
Differences in the sizes of the communities decrease the values of
balance, but affect mostly the average of all pairs, and not the
average of the elite pairs.

\begin{figure*}[tbp]
  \centering
  \includegraphics[width=0.9\textwidth]{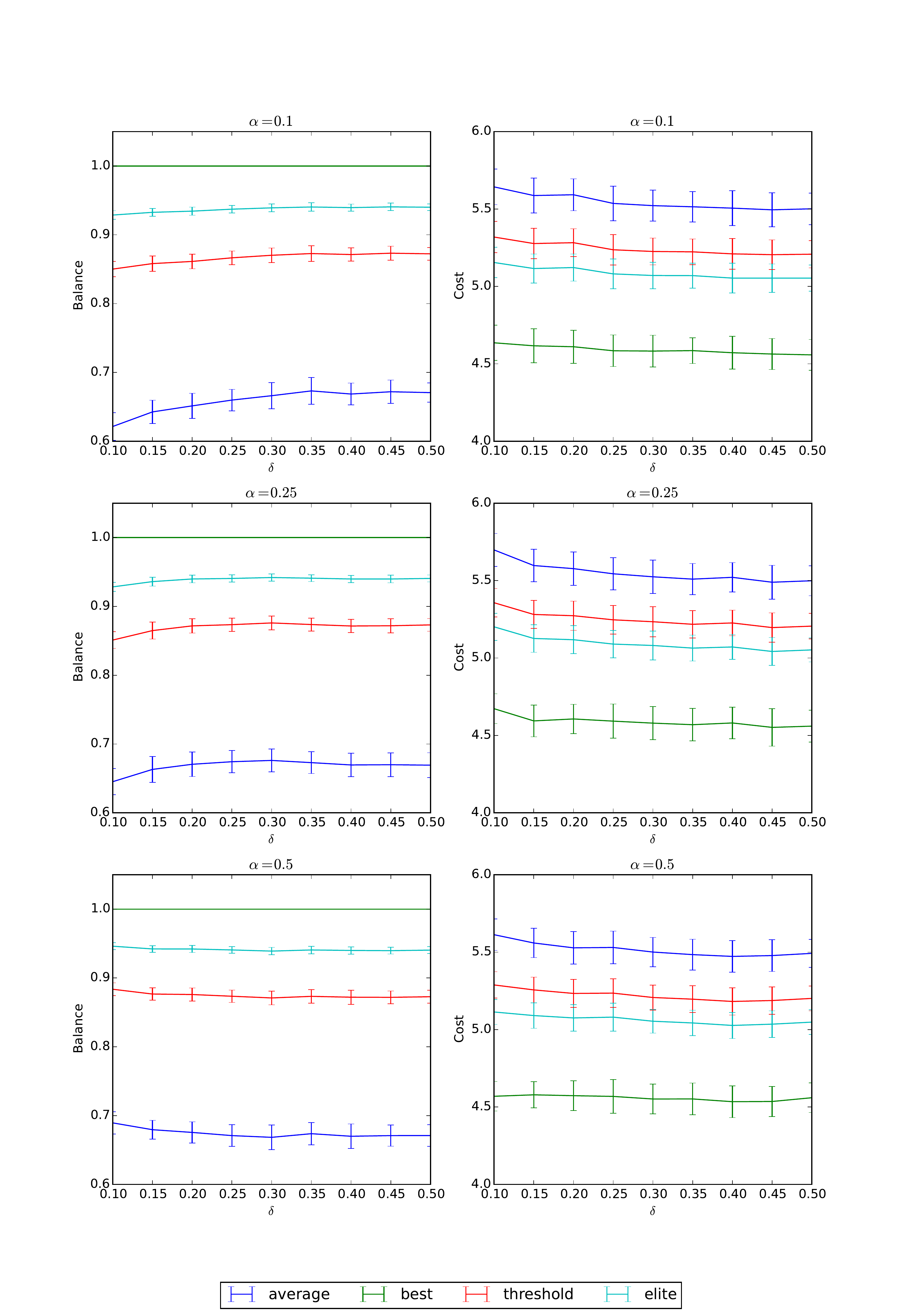}
  \caption{Effect of the community separation on balance and cost in a
    community model.  Community separation is controled by parameter
    $\delta$. Graphs are shown for different distributions of number
    of nodes for each community, as controled by parameter $\alpha$.
    Results are averages of 100~networks, each with 200~nodes; error
    bars show one standard deviation.}\label{fig:comm-delta}
\end{figure*}

Figure~\ref{fig:comm-delta} shows the effect of varying $\delta$ (for
some values of $\alpha$).  For communities of the same size
($\alpha=1/2$) there is almost no influence of $\delta$, with a
slightly better balance and worst average cost if $\delta$ is
small. For smaller values of $\alpha$ (i.e.\ if there is a larger
difference in the sizes of the two communities) a clear trend is seen
where smaller values of $\delta$ lead to worse values of balance and
cost.  This means that a network with communities of different sizes
and strong community separation is not well suited for this kind of
dynamics.

\begin{figure*}[tbp]
  \centering
  \includegraphics[width=0.9\textwidth]{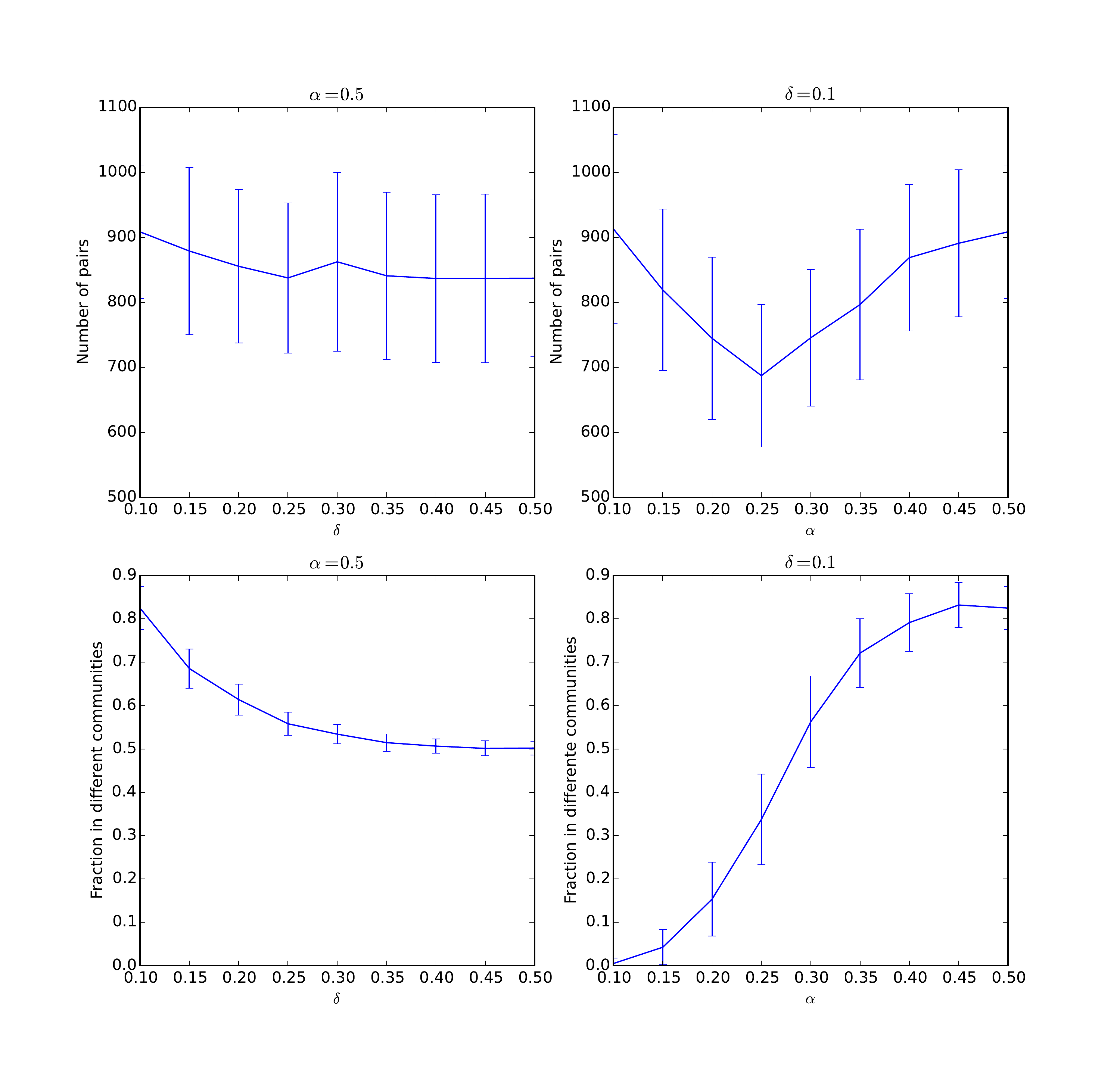}
  \caption{Number of pairs in the elite and fraction of those pairs
    that have a node in each community.  Results are shown fixing the
    number of nodes in each community through the parameter
    $\alpha = 1/2$ and varying the strength of connectivity between
    networks by changing $\delta$, or fixing $\delta=0.1$ and varying
    $\alpha$.}\label{fig:number-frac}
\end{figure*}

The previous results are complemented by the ones presented in
figure~\ref{fig:number-frac}, where we fix $\alpha=1/2$ and change
$\delta$ (left) or fix $\delta=0.1$ and change $\alpha$ (right), and
evaluate the number of pairs in the elite (top) and the fraction of
these elite pairs where each element is in a different community
(bottom).  On the top left we see that, for communities of the same
size, the number of pair in the elite is not affected by the strength
of the separation of communities.  The bottom left plot shows that for
small values of $\delta$ almost all pairs in the elite have nodes in
different communities.  This means that, under strong community
structure, a good efficiency can only be achieved by putting one
server in each community.  On the top right we see that in the case of
a relatively strong community structure ($\delta=0.1$), the number of
pairs in the elite decreases as $\alpha$ is decreased from $0.5$ to
$0.25$, but increases again afterward.  As $\alpha$ decreases, the
communities are of different sizes, and it becomes more difficult to
find pairs of nodes that at the same time are close to the client
nodes and equally divide those clients between themselves.  The
increase below $\alpha=0.25$ can be explained by looking at the bottom
right plot, where we see that fraction of elite pairs with nodes in
different communities sharply decreases as $\alpha$ decreases.  This
means that, as one of the communities decreases in size, it becomes
advantageous putting both server nodes in the largest community, as
the increased cost for the small community is of little total
influence.

\section{Conclusion}
\label{sec:concl}

This article has investigated the effect of distinct distribution of servers
in cloud computing environments with respect to three network topology,
namely ER, BA and modular.  In order to better discuss and organize
the investigation, we classified as elite the pairs of servers
with top performance regarding both communication cost and balance.

Several results have been obtained.  First, we have that ER generally
provides better balance in detriment of communication cost, while BA
provides complementary characteristics.  In addition, the elite pairs of
servers are more populous in the ER than in the BA networks, and the
difference between the best and average pairs is larger in the latter.
The investigation of the modular networks was performed while varying
the number of nodes in each community and the strength of connection
between them.  Though the balance is affected by the relative size of
the communities, little effect has been observed regarding communication
cost.  Also, for communities with similar size, the strength of interconnection
 between communities was not found to influence either communication 
 cost or balance.  However, if the communities have different sizes, less
 interconnection between them worsens both balance and cost.  When
 the separation between the communities is pronounced, most of the 
 elite pairs will have each of its server in different communities.
All in all, we have confirmed that the distribution of servers in cloud 
computing environments can be critical for the performance in terms
of communication cost and balance, with the best configurations 
depending heavily on the network topology.

Future works could address more than a pair of servers, other network
topologies, and consider the effect of specific network features on the 
performance.

\begin{acknowledgement}
  OMB is grateful to CNPq (307797/2014-7 and 484312/2013-8). LdFC is
  grateful to FAPESP (2011/50761-2), CAPES, NAP-PRP-USP, and CNPq
  (307333/2013-2),
\end{acknowledgement}

\bibliographystyle{unsrt}
\bibliography{cloud}

\end{document}